\documentclass[12pt]{article}
\usepackage{amsmath}
\usepackage{amssymb}
\usepackage{graphicx}
\newcommand{\be}{\begin{equation}}
\newcommand{\ee}{\end{equation}}
\newcommand{\ba}{\begin{eqnarray}}
\newcommand{\ea}{\end{eqnarray}}
\newcommand{\bc}{\begin{center}}
\newcommand{\ec}{\end{center}}
\newcommand{\ds}{\displaystyle}
\newcommand{\lra}{{\small $\leftrightarrow$}}
\begin{document}

\begin{center}
\bf  High-temperature spin polarization of high-mobility charge
carriers\\ in hybrid metal-semiconductor structures
\end{center}
\bigskip

\centerline{Meilikhov E.Z., Farzetdinova R.M.}
\medskip
\centerline{\small\it Kurchatov Institute, 123182 Moscow, Russia}
  \vspace{1cm}

\begin{center}
\begin{tabular}{p{15cm}} \footnotesize\quad  We consider magnetic
properties of the planar structure consisting of a ferromagnetic
metal, diluted magnetic semiconductor and the quantum well (by the
example of the hybrid heterostructure Fe--Ga(Mn)As--InGaAs). In the
framework of the mean-field theory, there is the significant
amplification of the ferromagnetism induced by the ferromagnetic
metal (Fe) in the system of magnetic impurities (Mn) due to their
indirect interaction via the conductivity channel in the quantum
well. As a result, the high-temperature ferromagnetism arises
leading to the spin polarization of  charge carriers (holes)
localized in the quantum well and preserving their high mobility.
\end{tabular}
\end{center}
\vspace{1cm}

\centerline{\bf Introduction}
\medskip

Among troublesome barriers on the way to the development of the
semiconductor spintronics, there are two principal ones -- the lack
of semiconductor materials and structures which would be (i)
ferromagnetic at high (room) temperature, and (ii) would possess
high enough mobility of charge carriers.  In this connection, there
may be promising hybrid ferromagnetic metal/semiconductor structures
whose magnetic properties are significantly determined by the
high-temperature ferromagnetism of the constituent metal~\cite{1,2},
and heterostructures based on diluted magnetic semiconductors with
removed doping and characterized by the high carrier mobility in the
quasi-two-dimensional conductivity channel~\cite{3}.

High-temperature magnetism in the semiconductor part of the first
type structures is conditioned by inducing magnetic order in
impurity atoms of the diluted magnetic semiconductor due to the
proximity effect~\cite{1}, however the mobility of polarized charge
carriers (holes) is very low at that ($\sim$1-10
cm$^2$/(V$\cdot$s)~\cite{4}). To remedy this, one could spatially
separate those atoms and holes to obtain the amplification of the
above-mentioned "seed" magnetic order by means of the indirect
interaction of impurity atoms through the "tails" of carriers' wave
functions mainly localized in the quantum well. It has been
experimentally shown that such a spatial separation results in a
high carrier mobility ($\sim10^3$ cm$^2$/(V$\cdot$s)~\cite{5}),
though obtained Curie temperatures (with no magnetic seed induced by
the proximity effect) do not exceed  250 K~\cite{6}.

In the present paper, we consider magnetic properties of hybrid
heterostructures (of
Fe--Ga$_{1-x}$Mn$_{x}$As--In$_y$Ga$_{1-y}$As-type, cf.~Fig. 1) where
one could realize both mentioned principles. High-mobility charge
carriers (holes) are concentrated in the two-dimensional quantum
well appearing in the narrow-band gap non-magnetic semiconductor
In$_y$Ga$_{1-y}$As near its junction with the wide-band gap magnetic
semiconductor Ga$_{1-x}$Mn$_{x}$As, whereas impurity atoms of the
latter are "magnetized" by Fe atoms. Fe film, at such, is in the
ferromagnetic single domain state being magnetized up to the
saturation (along the unit vector {\boldmath$ \rho$}$_0$ parallel to
the interface). Such a system, as we will show could combine
magnetic order with the high carrier mobility over a wide range of
temperatures.
\newpage
\centerline{\bf Mean-field model}
\medskip

The magnetization of the diluted magnetic semiconductor (which is
parallel to the interface due to the shape magnetic anisotropy) is
significantly non-uniform along the growth axis $z$ (cf.~Fig. 1). It
could be conveniently characterized by the local magnetization
$-1\leqslant j({h})\equiv M({h})/M_s\eqslantless1$, directed along
the unit vector {\boldmath$ \rho$}$_0$ ($M(h)$ is the local
magnetization at the distance $h$ from the
Fe/Ga$_{1-x}$Mn$_{x}$-interface plane, $M_s$ is the saturation
magnetization). In the framework of the mean-field theory it is
defined by the equation
 \be\label{1}
j(h)=B_S\left[-\frac{W(h)}{kT}\right],
  \ee
where $B_S$ is the Brillouin function for the spin $S$ of Mn atoms,
 \be\label{2}
W(h)= \sum\limits_i w_{\rm Fe}({\bf R}_{i})+\sum\limits_k w_{\rm
Mn}({\bf r}_{k})
  \ee
is the energy of the magnetic interaction of a given Mn atom with
other parts of the structure. This energy is the sum of energies
$w_{\rm Fe}$ and $w_{\rm Mn}$ of its pair interactions with
individual Fe and Mn atoms, spaced at distances ${\bf R}_i$ and
${\bf r}_k$ from the specified Mn atom, respectively. Summation is
performed over all Fe atoms (in the first sum) and all Mn atoms (in
the second sum).

When calculating the first sum in (\ref{2}),  the interaction of Mn
atoms will be considered being antiferromagnetic one~\cite{1} and
corresponding effective magnetic field being also directed along the
unit vector {\boldmath$ \rho$}$_0$. Because considered Fe and Mn
atoms are located in media characterized, in general, by different
lengthes of the exchange interaction, the result of that interaction
is not merely the function of the distance $R_i$ between those
atoms. However, for not to complicate calculations with
non-principal details we will proceed below from the following model
spatial dependence
 \be\label{3}
w_{\rm Fe}(R_{i})=-J_{\rm Fe}\exp[-R_i/\ell_{\rm
Fe}]\mbox{{\boldmath$\rho$}$_0\cdot {\bf S}_i$},
  \ee
where $J_{\rm Fe}$ and $\ell_{\rm Fe}$ are, correspondingly, the
characteristic energy and length of the considered interaction for
Mn atom with the spin ${\bf S}_i$. Putting the coordinate origin in
the interface plane, we assume that Fe layer occupies the interval
$-L_{\rm Fe}<z<0$, and the semiconductor film Ga(Mn)As takes up the
range $0<z<L_{\rm Mn}$ (cf. Fig. 1). Then, in the continual
approximation
 \be\label{4}
 \sum\limits_i w_{\rm Fe}(R_{i})= -J_{\rm Fe}\,Sn_{\rm
Fe}\!\!\!\int\limits_{z=-L_{\rm Fe}}^0\int\limits_{\rho=0}^\infty
\exp\left[-\sqrt{\rho^2+(h-z)^2}/\ell_{\rm Fe}\right]\!2\pi\rho
d\rho dz,
 \ee
where $n_{\rm Fe}$ is the concentration of Fe atoms. From
Eq.~(\ref{4}) it follows
 \be\label{5}
 \sum\limits_i w_{\rm Fe}(R_{i})=- 4\pi n_{\rm
Fe}\ell_{\rm Fe}^3SJ_{\rm Fe}F(h),\quad F(h)=e^{-h/\ell_{\rm
Fe}}\left[1+\frac{h}{2\ell_{\rm Fe}}-\left( 1+\frac{L_{\rm
Fe}+h}{2\ell_{\rm Fe}}\right)e^{-L_{\rm Fe}/\ell_{\rm Fe}} \right].
 \ee

The calculation of the second sum in (\ref{2}) should be preceded by
the following comment. Presently, there is no full understanding the
nature of the ferromagnetism in diluted magnetic semiconductors.
Among mechanisms leading to the ferromagnetic ordering of magnetic
impurities' spins they consider different types of their indirect
interaction via mobile charge carriers: RKKY-exchange~\cite{7},
kinematic exchange~\cite{8,9}, etc.~\cite{8}. In addition, there is
the universal Bloembergen-Rowland mechanism~\cite{10}, that does not
require the existence of charge carriers (or their high
concentration resulting in the carrier degeneracy) and could be drew
for explaining ferromagnetism in the systems like Ga(Mn)As,
Ga(Mn)N~\cite{11}. In this connection, we will use the RKKY
mechanism as a \emph{model} interaction (without insisting upon its
universality for the considered systems).

In the \emph{really two-dimensional} case, when magnetic impurities
are situated in the plane of the two-dimensional gas of charge
carriers, the energy of their RKKY interaction equals
 \be\label{2D}
  w_{\rm Mn}^0(\rho_k)=-J_{\rm Mn}\phi(\rho_{k}){\bf S}_0\cdot {\bf S}_k,
 \ee
where $\rho_{k}$ is the distance between a given Mn atom (with the
spin ${\bf S}_0$) and any one of other Mn atoms (with the spin ${\bf
S}_k$). The characteristic interaction energy  $J_{\rm Mn}$ and the
function $\phi(r_{ik})$ define by the relationships~\cite{12}
 \be\label{6}
J_{\rm Mn}=\left(J_{pd}^2\frac{m^*a_0^2}{4\pi\hbar^2}\right),\quad
\phi(\rho_{k})=(k_F
a_0)^2\left[J_0(k_F\rho_{k})N_0(k_F\rho_{k})+J_1(k_F\rho_{k})N_1(k_F\rho_{k})\right]e^{-\rho_{k}/\ell_{\rm
Mn}},
 \ee
where   $J_{pd}$ is the energy of $p$-$d$ interaction, $k_F=(\pi
N_s)^{1/2}$ is the Fermi wave number of carriers, $J_n$ and $N_n$
are Bessel functions (with $J_{\rm Mn}\approx0.1$ eV~\cite{13}),
$a_0\approx6$\AA\, is the side of GaAs cubic cell. The exponential
factor in Eq. (\ref{6}) takes into account the damping of the
interaction (with the characteristic length $\ell_{\rm Mn}$) due to
the scattering of carriers~\cite{14}. In the \emph{bulk} diluted
magnetic semiconductor Ga$_{1-x}$Mn$_x$As with the actual impurity
concentration $x\approx0.05$, the hole mobility is rather low:
$\mu_h$=1-10 cm$^2$/V$\cdot$s~\cite{4}, that corresponds to their
mean free path along the quantum well $\ell_0\sim a_0$.

Those relations fall into the case when Mn atoms and charge
carriers, providing their interaction,  are placed together within
the quasi-two-dimensional conductivity channel.

In our case, mobile charge carriers are localized in the quantum
well being spatially separated from Mn atoms. The interaction of
magnetic atoms occurs due to the leakage of the carrier wave
function into the region of their  arrangement. Then, as it has been
shown in~\cite{3}, the three-link "chain"  of interactions works:
(i)~interaction of a given impurity $i$ with the channel + (ii)
"transfer" of the interaction along the quasi-two-dimensional
channel + (iii) interaction of a removed impurity $k$ with the
channel. Each of them leads to the specific factor in the total
expression for the interaction energy:
 \be\label{7}
 w_{\rm Mn}(\rho_{k},
z_k)= w_{\rm Mn}^0(\rho_k)\cdot \left[\frac{\psi^2_b(z_i)}{\psi_{\rm
max}^2}\right]\cdot \left[\frac{\langle\psi_a^2\rangle}{\psi_{\rm
max}^2}\right]\cdot \left[\frac{\psi^2_b(z_k)}{\psi_{\rm max}^2}\right].
 \ee
Here, $\psi_{\rm max}$ is the maximum value of the carrier wave
function in the well (corresponding to the peak of their
concentration in the quasi-two-dimensional channel), $\psi_b(z)$ is
their wave function, "leaking" into the magnetic semiconductor, and
$\langle\psi_a^2\rangle$ is the average value of squared wave
function in the well that could be found by means of the
relation~\cite{3}
 \be\label{average}
\langle\psi_a^2\rangle=\frac{1}{\sqrt{3}}\left(\frac{1}{2a}\int\limits_{L_{\rm
Mn}}^\infty\psi_a^2(z)dz\right),
 \ee
where
$$
  a=\left\{\,\int\limits_{L_{\rm Mn}}^\infty z^2\psi_a^2(z)dz\Big/\int\limits_{L_{\rm Mn}}^\infty
\psi_a^2(z)dz-\left[\,\,\int\limits_{L_{\rm Mn}}^\infty
z\psi_a^2(z)dz\Big/\int\limits_{L_{\rm Mn}}^\infty
\psi_a^2(z)dz\right]^2\right\}^{1/2}
$$
is the effective half-width of the wave function in the well.

Spatial separation of impurities and charge carriers in the
considered structures leads, on the one hand, to weakening the
indirect interaction, and results, on the other hand, in increasing
the mean free path of carriers that promotes the strengthening of
that interaction. If magnetic impurities are situated out of the
well, the mean carrier free path $\ell$ increases\cite{16}
(comparing to its value $\ell_0$ corresponding to the case when they
are inside the well):
 \be\label{ell}
\ell_{\rm Mn}=\ell_0\frac{\langle\psi_a^2\rangle}{\langle\psi_b^2\rangle },
 \ee
where $\langle\psi_b^2\rangle=(1/2\sqrt{3}\,b)\int_0^{L_{\rm
Mn}}\psi_b^2(z)dz$ is the averaged (over the impurity layer) value
of the squared wave function with the effective half-width
$$
b=\left\{\,\int\limits_0^{L_{\rm Mn}}
z^2\psi_a^2(z)dz\Big/\int\limits_0^{L_{\rm Mn}}
\psi_a^2(z)dz-\left[\,\,\int\limits_0^{L_{\rm Mn}}
z\psi_a^2(z)dz\Big/\int\limits_0^{L_{\rm Mn}}
\psi_a^2(z)dz\right]^2\right\}^{1/2}.
$$

The respective increase of the carrier mobility could be very
significant. For instance, in~\cite{5} it has amounted to 2-3 orders
of value that has allowed Shubnikov oscillations of the conductivity
and quantum Hall effect in the two-dimensional channel of the
single-well structure with the concentration of removed (from the
channel) impurities $x\approx0.05$.

Thus, the removed doping promotes increasing the carrier mobility
and, as a consequence, leads to increasing the energy of indirect
interaction of magnetic impurities.

Introducing the cylindrical coordinate system with the former origin
of the $z$-axis and the radius-vector~{\boldmath$\rho$}, parallel to
the interface plane,  we will characterize the magnetic order
arising in the system of impurities by the reduced magnetization
$-1<j=j(z,r)<1$, which coincides with the impurity spin polarization
degree. Neglecting the crystal and surface anisotropy, one notices
that the shape anisotropy and the system symmetry result in that the
local magnetization is  everywhere directed along the plane of the
impurity layer and depends on~$z$ only: $j=j(z)$.

In the continual approximation, the total energy of the indirect
interaction of the impurity, located in the point $z=h$, $\rho=0$,
with all surrounding impurities equals
 \ba\label{8}
\sum\limits_k w_{\rm Mn}({\rho}_{k}, z_k)\approx\int\!\!\!\int w_{\rm Mn}(\rho,
z)2\pi\rho d\rho dz=\hspace{70pt}\nonumber\\
= - 2\pi n_{\rm Mn}S^2 J_{\rm
Mn}\left[\frac{\langle\psi_a^2\rangle}{\psi_{\rm
max}^2}\right]\left[\frac{\psi^2_b(h)}{\psi_{\rm
max}^2}\right]\int\limits_{\rho_{\rm min}}^\infty\phi(\rho)\rho
d\rho\cdot\!\!\int\limits_{z=0}^{L_{\rm Mn}}
\left[\frac{\psi^2_b(z)}{\psi_{\rm max}^2}\right]j(z)dz,
 \ea
where it has been taken into account that the distance between
impurities could not be smaller than the certain  minimum distance,
which for Mn atoms substituting Ga atoms in GaAs lattice equals
$\rho_{\rm min}=a_0/\sqrt{2}$.

Let us introduce the function
 \be\label{9}
 \Phi(k_F,\ell)=-\frac{\pi}{a_0^2}\int\limits_{\rho_{\rm min}}^\infty
\phi(\rho)\rho\, d\rho,
 \ee
Then Eq. (\ref{8}) becomes
 \be\label{10}
\sum\limits_k w_{\rm Mn}({\rho}_{k}, z_k)\approx2n_{\rm Mn} a_0^3
S^2J_{\rm
Mn}\Phi(k_F,\ell)\left[\frac{{\langle\psi_a^2\rangle}}{{\psi_{\rm
max}^2}}\right]\left[\frac{ {\psi^2_b(h)}}{{\psi_{\rm
max}^2}}\right]\left[\frac{1}{a_0}\int\limits_{z=0}^{L_{\rm Mn}}
\!\!\!\!\left[{\psi^2_b(z)}/{\psi_{\rm max}^2}\right]j(z)dz\right].
 \ee

Substituting (\ref{5}) and (\ref{10}) in Eq.  (\ref{1}), one comes
to the equation defining the spatial magnetization in the magnetic
semiconductor:
 \be\label{11}
j(h)=B_S\!\!\left[\frac{\mu H(h)}{kT}\right]=B_S\left[\frac{C_{\rm
Fe}(h)}{\tau}+\frac{C_{\rm
Mn}(h)}{\tau}\left(\frac{1}{a_0}\int\limits_{z=0}^{L_{\rm Mn}}\!
\left[{\psi^2_b(z)}/{\psi_{\rm max}^2}\right]j(z)dz\right)\right],
 \ee
where $\tau=kT/J_{\rm Mn}$ is the reduced temperature, $C_{\rm
Fe}(h)=4\pi n_{\rm Fe}a_0^3S(J_{\rm Fe}/J_{\rm Mn})(\ell_{\rm
Fe}/a_0)^3F(h)$, $C_{\rm Mn}(h)=C_0 \left[{\psi^2_b(h)}/{\psi_{\rm
max}^2}\right]$,\, $C_0= 2 n_{\rm Mn} a_0^3S^2\Phi(k_F,\ell_{\rm
Mn})\left[{\langle\psi_a^2\rangle}/{\psi_{\rm max}^2}\right]$.

The solution of that equation is
 \be\label{12}
j(h)=B_S\left[\frac{C_{\rm Fe}(h)}{\tau} +\gamma(\tau) C_{\rm Mn}(h)\right],
 \ee
where the parameter $\gamma(\tau)$ is defined self-consistently by
substituting the function (\ref{12}) in the relation (\ref{11}).

Let us consider, firstly, the case $n_{\rm Fe}=0$, i.e. the system
without the ferromagnetic Fe layer, but with the indirect
interaction of magnetic impurities via charge carriers in the
quantum well~\cite{3}. Then $C_{\rm Fe}(h)\equiv0$ and the stated
substitution leads to the equation
 \be\label{13}
\gamma=\frac{1}{\tau}\left(\frac{1}{a_0}\int\limits_{z=0}^{L_{\rm
Mn}}\left[{\psi^2_b(z)}/{\psi_{\rm max}^2}\right]B_S\left[\gamma
C_{\rm Mn}(z)\right]dz\right),
 \ee
which determines the parameter $\gamma$. It has nonzero solution in
the low-temperature region $\tau<\tau_{\rm C}$ only, where
$\tau_{\rm C}$ is the Curie temperature.

Near the Curie temperature, the magnetization is low ($j\to 0$).
According to (\ref{12}) it is possible at $\gamma\to 0$ only. Using
the series $B_S(x)=b_1x-b_3x^3+\ldots$ with $b_1=(S+1)/3S$, $b_3>0$,
one finds the solution of Eq. (\ref{13})
 \be\label{14}
\gamma^2=\frac{b_1}{b_3C_0^2}\left(\frac{1}{a_0}\int\limits_{z=0}^{L_{\rm
Mn}}\left[{\psi^4_b(z)}/{\psi_{\rm
max}^4}\right]dz-\frac{\tau}{b_1C_0}\right)
\left(\frac{1}{a_0}\int\limits_{z=0}^{L_{\rm
Mn}}\left[{\psi^8_b(z)}/{\psi_{\rm max}^8}\right]dz\right)^{-1},
 \ee
which at $\gamma=0$ provides the Curie temperature:
 \be\label{15}
\tau_{\rm C}=b_1C_0\left(\frac{1}{a_0}\!\int\limits_{z=0}^{L_{\rm
Mn}}\!\!\!\!\left[{\psi^4_b(z)}/{\psi_{\rm max}^4}\right]dz\right)=\tau_0
\left(\frac{1}{a_0}\!\int\limits_{z=0}^{L_{\rm
Mn}}\!\!\!\!\left[{\psi^4_b(z)}/{\psi_{\rm
max}^4}\right]dz\right)\left[\langle\psi_a^2\rangle/\psi^2_{\rm max}\right].
 \ee
where $\tau_0=(2/3)\,n_{\rm Mn} a_0^3\,S(S+1)\Phi(k_F,\ell_{\rm
Mn})$. It determines the temperature range for existing "intrinsic"
(not induced by Fe film) ferromagnetism in Ga(Mn)As.

Assuming now $n_{\rm Fe}\ne0$ but $\ell_{\rm Mn}=0$, we come to the
situation with no indirect interaction of magnetic impurities. In
that case, impurities are magnetized due to the exchange interaction
with the magnetically ordered system of Fe atoms only. The spatial
distribution of such an induced magnetization is defined by
Eq.~(\ref{12}), where one should assume $\gamma=0$. One could
estimate the width $L_{\rm ind}$ of that distribution (at the level
$1/2$) from the condition $B_S[C_{\rm Fe}(h)/\tau]=1/2$, which for
$S=5/2$~\cite{15} gives $C_{\rm Fe}(h)/\tau\approx1.4$. At $n_{\rm
Fe}a_0^3\approx2$, $J_{\rm Fe}/J_{\rm Mn}\sim1$, $\ell_{\rm Fe}\sim
a_0$, $\tau\sim\tau_{\rm C}$ one finds $L_{\rm ind}\approx
4\ell_{\rm Fe}$. Thus, the induced magnetization of Mn atoms exists
in a thin layer of the thickness $L_{\rm ind}\sim 20{\rm\AA}$ near
the  Fe/Mn interface. It is just that layer which serves as a
magnetic "seed" being amplified due to the indirect inter-impurity
interaction.

At last, in the general case the spatial distribution of the
magnetization is defined by Eq.~(\ref{12}), where the parameter
$\gamma(\tau)$ is the root of the equation
 \be\label{16}
\frac{1}{a_0}\int\limits_{z=0}^{L_{\rm Mn}}\left[{\psi^2_b(z)}/{\psi_{\rm
max}^2}\right]B_S\left[\frac{C_{\rm Fe}(z)}{\tau}+\gamma(\tau) C_{\rm
Mn}(z)\right]dz-\gamma(\tau)\tau=0.
 \ee

Formulae (\ref{12}), (\ref{16}) is the main result of the present
work. To determine the magnetization with the help of those
relations, it is necessary to find the wave functions  $\psi_a(z)$,
$\psi_b(z)$ of charge carriers in different parts of the considered
structure.

Such a problem for the heterostructure being the contact of two
different semiconductors (e.g., GaAs and GaInAs), one of which
(GaAs) is diluted by Mn atoms, has been considered by us
early~\cite{3}. Opposite in sign charges of ionized impurities and
mobile charge carriers in the well produce the electric field
{\Large$\varepsilon$}, which is directed along the normal ($z$-axis)
to the heterojunction plane and makes the potential to be
non-uniform: $U=U(z)$. The exact self-consistent determination of
the potential $U(z)$ and wave functions of mobile charge carriers
requires the consistent solution of Schr\"{o}dinger and Poisson
equations that is usually found by  numerical iterative
methods~\cite{16} (relevant calculations could be performed, for
example, by means of the openly accessible package~\cite{17}).
However, our aim is to derive simple \emph{analytical} expressions
describing magnetic properties of the considered system. Thereby, we
will use the heterojunction model with the triangle well bottom and
barrier top. Tests indicate that wave functions found with that
"triangle" model are very close to "exact" results~\cite{3}.
Respective expression for the carrier potential energy has the form
 \be\label{17}
U(z)=\left\{
\begin{tabular}{ll}
$U_0+(z-L_{\rm Mn})e${\Large$\varepsilon$} ,&$z<L_{\rm Mn},$\\
$(z-L_{\rm Mn})e${\Large$\varepsilon$} ,&$z>L_{\rm Mn}$\\
\end{tabular}
\right.
 \ee
(as previously, $z=L_{\rm Mn}$ corresponds to the heterojunction
plane, magnetic impurities are situated in the region $0<z<L_{\rm
Mn}$). The slope of the well bottom is determined by the electric
field {\Large$\varepsilon$}$\approx(4\pi/\kappa_0)eN_s$, produced by
charges located in the well.

As before, we assume the carrier density being not too high, so that
the lowest energy level occurs to be populated only, and the
effective width of the well being so small that mixing of light and
heavy hole subbands could be neglected. Near the heterojunction
($z=L_{\rm Mn}$) the wave function of carriers on the lowest energy
level $E$ has the form~\cite{3}
 \be\label{18}
  \psi(z)=\mathbb C\left\{
\begin{tabular}{ll}
$\psi_b(z)\equiv{\rm Ai}[q(z-L_{\rm Mn})-\varepsilon/q^2 a_0^2]$,&$z<L_{\rm Mn}$\\
&\\
$\psi_a(z)\equiv\frac{\ds{\rm Ai}(-\varepsilon/q^2a_0^2)}{\ds{\rm
Bi}(-\varepsilon/q^2a_0^2+u/q^2a_0^2)\rule{0pt}{12pt}}\,
{\rm Bi}[q(z-L_{\rm Mn})-\varepsilon/q^2 a_0^2+u/q^2a_0^2]$,&$z>L_{\rm Mn}$\\
\end{tabular}
\right.,
 \ee
where ${\rm Ai}(z)$,  ${\rm Bi}(z)$ are Airy functions,
 \be\label{19}
u\equiv 2m^*a_0^2U_0/\hbar^2,\quad\varepsilon\equiv 2m^*a_0^2E/\hbar^2, \quad
q\equiv\left(\frac{2m^*}{\hbar^2}e\mbox{\Large$\varepsilon$}\right)^{1/3},
 \ee
 $\varepsilon$ is the reduced energy of the populated level in the
 well defined by the equation
 \be\label{20}
{\rm Ai}(-\varepsilon/q^2a_0^2){\rm Bi}'(-\varepsilon/q^2
a_0^2+u/q^2a_0^2)-{\rm Ai}'(-\varepsilon/q^2a_0^2){\rm Bi}(-\varepsilon/q^2
a_0^2+u/q^2a_0^2)=0,
 \ee
coefficient $\mathbb C$ should be found from the normalization
condition. \vspace{0.5cm}

\centerline{\bf Results}
\medskip

Though above we had to do with the concrete structure
Fe--Ga(Mn)As--In(Ga)As, the qualitative character of our
consideration makes using accurate values of those parameters that
govern its behavior to be excessive. Therefore, we assume $n_{\rm
Fe}a_0^3=2$, $n_{\rm Mn}a_0^3=0.15$ (that corresponds to
$x\approx0.1$), $J_{\rm Fe}/J_{\rm Mn}=1$, and for other parameters
we suggest typical values $L_{\rm Fe}=L_{\rm Mn}=7a_0$~\cite{1},
$\ell_{\rm Fe}=0.75a_0$, $\ell_0=3a_0$ (that for the bulk diluted
magnetic semiconductor Ga(Mn)As corresponds to the hole mobility
$\sim10$ cm$^2$/V$\cdot$s).

The temperature range of existing "intrinsic" (not induced by Fe
film) ferromagnetism in Ga(Mn)As, found for that set of parameters,
is bounded from above by a rather low Curie temperature $\tau_{\rm
C}\approx 0.026$ ($T_{\rm C}\approx25$~K). In Fig.~2, the spatial
distribution $j_{\rm Mn}(z)$ of the intrinsic local magnetization of
Mn atoms is shown for the temperature $\tau$, close the critical one
(the curve Mn\lra Mn). In the same figure, spatial distributions of
Mn magnetization, induced by the exchange interaction with Fe atoms
and decaying with moving off the interface Fe/Ga(Mn)As ($z=0$), are
represented (curves Fe\lra Mn) for the case when their indirect
interaction is "switched off". At last, curves (Fe\lra Mn+Mn\lra Mn)
are the result of the combined action of the two magnetic ordering
mechanisms revealing good shows of the induced ferromagnetism
amplification due to the indirect interaction of magnetic Mn
impurities. Remarkably, significant amplification of the induced
magnetization keeps at temperatures which are higher than the Curie
temperature corresponding to the intrinsic ferromagnetism of
Ga(Mn)As.

Magnetization {$j_{\rm Mn}(z=L_{\rm Mn})$} near the heterojunction
plane is of special interest because it is just this value
determines the spin polarization degree of charge carriers in
two-dimensional conductivity channel. Let $N_s^-$, $N_s^+$ be
concentrations of two-dimensional holes with spins antiparallel and
parallel to the magnetization, respectively. Then, the spin
polarization degree $\xi=(N_s^- - N_s^+)/N_s$ ($N_s=N_s^- +N_s^+$ is
their total concentration) of holes is defined by the effective
"magnetic" spin-dependent potential which for the bulk diluted
magnetic semiconductor with the uniform magnetization  $j_{\rm Mn}$
has the form  $V_{\rm mag}=n_{\rm Mn}a_0^3J_{pd}\,\sigma S_{\rm
Mn}j_{\rm Mn}$~\cite{18}, where $\sigma=\pm 1/2$ is the hole spin.
In the considered case, when charge carriers and magnetized Mn atoms
are spatially separated, that relation should be added by the factor
allowing for the fact that their interaction occurs through the tail
of the carrier wave function which, additionally, is non-uniform
within the channel region:
 \be\label{21}
\langle V_{\rm mag}\rangle=j_{\rm Mn}(L_{\rm Mn})\,n_{\rm
Mn}a_0^3J_{pd}\,\sigma S_{\rm Mn}\cdot \left[\psi^2_a(L_{\rm
Mn})/\langle\psi_a^2\rangle\right].
 \ee

The magnetic potential (\ref{21}) leads to splitting the energy
level $E$ in two spin sub-levels with energies $E^+=E+V_{\rm mag}$
and $E^-=E-V_{\rm mag}$. Concentrations of two-dimensional carriers
at each of them are defined by relations $N_s^\pm\propto E^\pm$,
wherefrom it follows
 \be\label{22}
\xi=2\langle V_{\rm mag}\rangle/E= j_{\rm Mn}(L_{\rm Mn})\cdot\xi_0,
 \ee
where $\xi_0=2n_{\rm Mn}a_0^3\,\sigma S_{\rm Mn}(J_{pd}/E)
\left[\psi^2_a(L_{\rm Mn})/\langle\psi_a^2\rangle\right]$. In the
considered system $\xi_0\sim0.5$, so that $\xi\sim j_{\rm Mn}(L_{\rm
Mn})$. The amplification of the magnetization in the region
adjoining the heterojunction results in the proportionate increasing
of the spin polarization of charge carriers in the two-dimensional
channel.

Fig.~3 demonstrates temperature dependencies of the magnetization of
Mn atoms near the heterojunction plane: the lower curve (Fe\lra Mn)
shows the induced magnetization, the upper curve (Fe\lra Mn+Mn\lra
Mn) -- the induced one, amplifying by the indirect interaction. The
latter corresponds also (on a certain scale) to the temperature
dependence of the  spin polarization in the hole channel. Evident
magnetization $j_{\rm Mn}(L_{\rm Mn})$ ($\sim$10\%) remains up to
temperatures $\tau\sim20\tau_{\rm C}\sim0.5$, that corresponds to
$T\sim 500$ K.

In the insert, the temperature dependence of the respective
"amplification factor" $K_j$, equal to the ratio of the two
mentioned magnetizations, is shown. The maximum amplification occurs
at $\tau\gg\tau_C$  and comes about $K_j\approx1.6$. Though this
effect, as such, could be important, but the most interesting
feature of the considered structure is the significant mobility
increasing of spin-polarized charge carriers (in two-dimensional
conductivity channel): in comparison with the mobility in the bulk
Ga(Mn)As,  it increases  according to (\ref{ell}) by $\ell_{\rm
Mn}/\ell_0\approx 25$ times and reaches the value $\mu_h\sim10^3$
cm$^2$/V$\cdot$s for the accepted parameters' set.

\vspace{0.5cm}\centerline{\bf Conclusions} \medskip

Magnetic properties of the planar structure Fe--Ga(Mn)As--In(Ga)As,
which consists of the diluted magnetic semiconductor bordering upon
the ferromagnetic metal (on one side) and  upon the quantum well (on
another side), are considered. In the framework of the mean-field
theory, there has been demonstrated the significant amplification of
the magnetization, induced by the ferromagnetic metal, in the
semiconductor region close to the interface due to the indirect
interaction of magnetic impurities via the conductivity channel.
Existing  evident high-temperature  magnetization in the considered
structure (and, hence, the noticeable spin polarization of carriers,
too) is provided by the interaction of Mn atoms with Fe film
(keeping magnetization up to $\sim$1000~K), and the high mobility of
spin-polarized charge carriers -- by their moving from the charged
impurities. Such a favorable combination of the two important
parameters in the considered structures holds out a hope that they
could be of interest as possible elements of different spintronic
devices.

\newpage
\newpage\centerline{\bf Figure captions}
\bigskip

Fig. 1. Heterostructure
Fe--Ga$_{1-x}$Mn$_{x}$As--In$_y$Ga$_{1-y}$As.
2D -- quasi-two-dimensional conduc\-ti\-vity channel in the quantum well.\\

Fig. 2. Spatial distributions of Mn atoms' magnetization.
 (Fe\lra Mn) --
induced magnetization, (Fe\lra Mn+Mn\lra Mn) -- induced
magnetization amplified by the indirect interaction, (Mn\lra Mn) --
intrinsic magnetization. Structure parameters: $u=0.36$,
$qa_0=0.25$, $k_Fa_0=0.1$, $L_{\rm Fe}=L_{\rm Mn}=7a_0$,
$\ell_0=3a_0$, $n_{\rm Fe}a_0^3=2$, $n_{\rm Mn}a_0^3=0.15$,
$\ell_{\rm Fe}=0.75a_0$, $J_{\rm Fe}/J_{\rm Mn}=1$.\\

Fig. 3.  Temperature dependencies of Mn atoms' magnetization near
the heterojunction plane: (Fe\lra Mn) -- induced magnetization,
(Fe\lra Mn+Mn\lra Mn) -- induced magnetization amplified by the
indirect interaction. Stracture parameters are the same as in Fig.
2. In the inset: temperature dependence of the amplification factor.\\

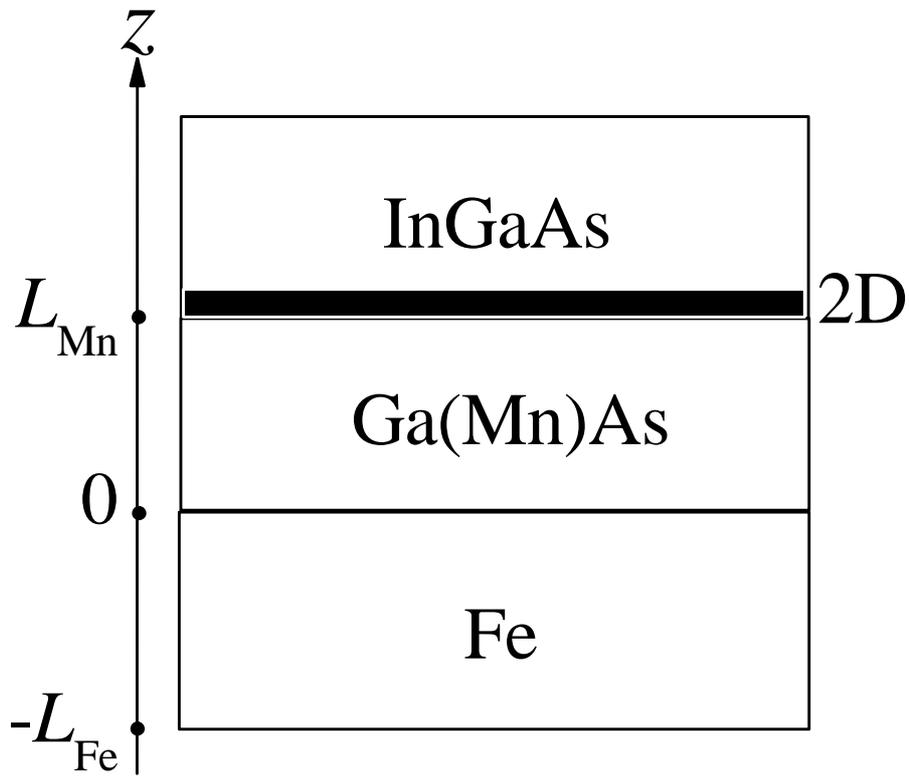
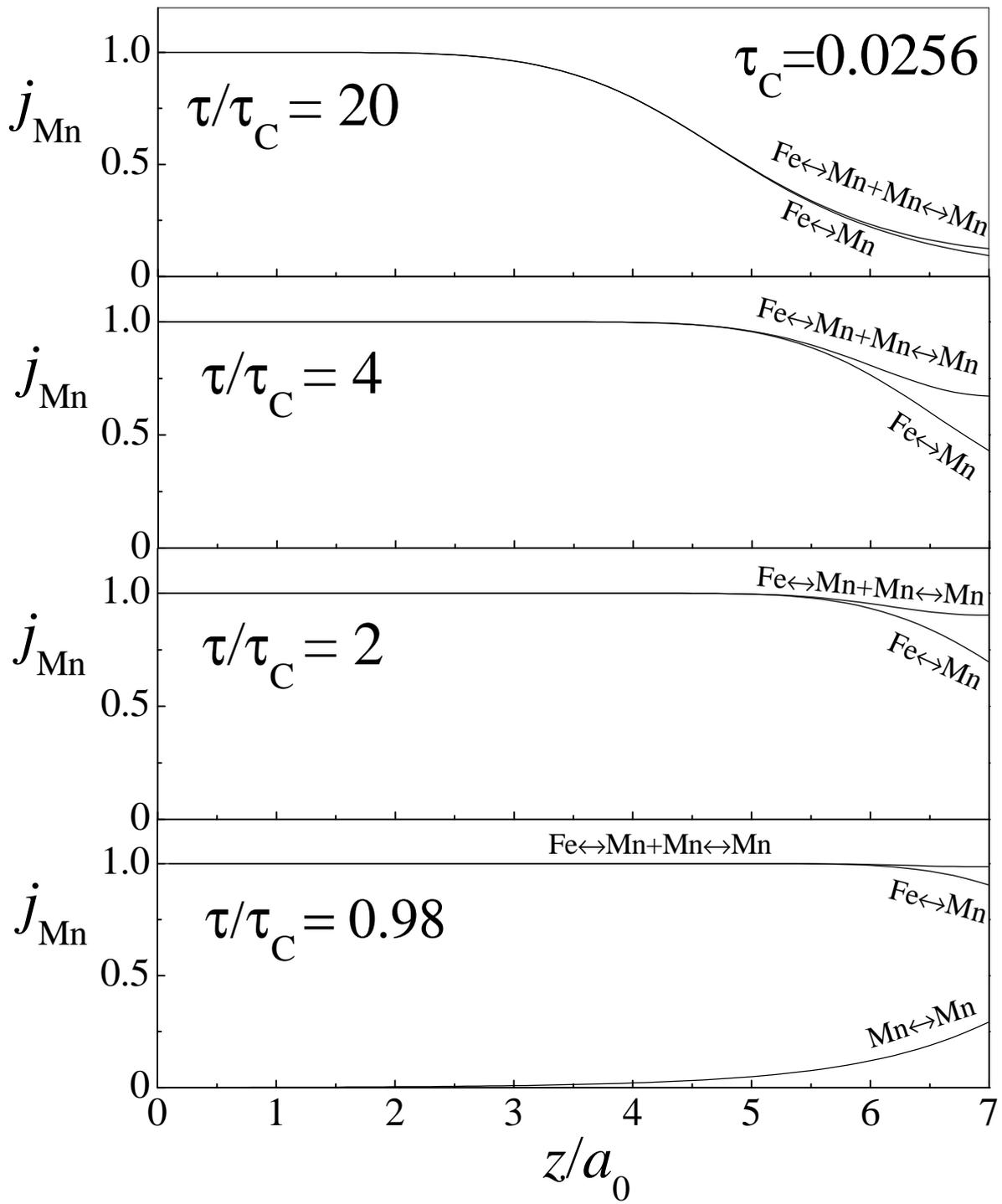
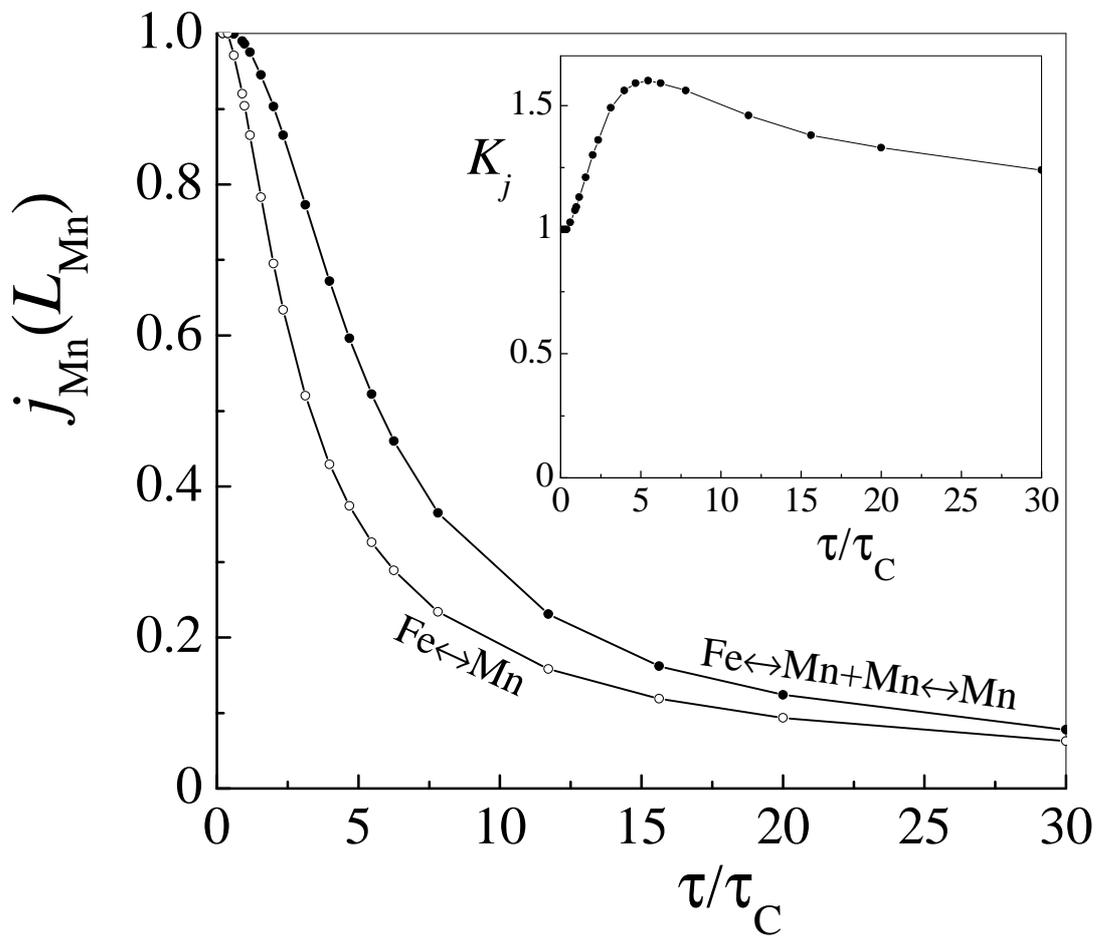
\end{document}